\documentclass[11pt,twoside]{article}  
\usepackage{adassconf}
\usepackage{asp2006}

\begin{document}   

%
%
%

\paperID{A.10 }

%
%
%
%

\title{A new model for archiving synoptic data in the VISTA Data Flow System.}
\titlemark{Archiving Synoptic Data in VDFS.}

%
%
%

\author{Nicholas Cross, Ross Collins, Nigel Hambly, Mike Read and Eckhard Sutorius}
\affil{Scottish Universities Physics Alliance, Institute for Astronomy, University of Edinburgh, Blackford Hill, Edinburgh, EH9 3HJ, U.K.}

%
%

\contact{Nicholas Cross }
\email{njc@roe.ac.uk}

%
%
%
%
%

\paindex{Cross, N.}
\aindex{Collins, R.}
\aindex{Hambly, N.}
\aindex{Read, M.}
\aindex{Sutorius, E.}

%
%

\authormark{Cross et al. }

%
%

\keywords{archives, astronomy!surveys, databases!design, Python,
WFCAM!Science Archive, VISTA!Science Archive, surveys!data products,
synoptic, variables}


\begin{abstract}
The VISTA Data Flow System comprises nightly pipeline and archiving of
near infrared data from UKIRT-WFCAM and VISTA. This includes multi-epoch 
data which can be used to find moving and variable objects. 
We have developed a new model for archiving these data which gives the 
user an extremely flexible and reliable data set that is easy to query through
an SQL interface. We have introduced several new database tables into our 
schema for deep/synoptic datasets. We have also developed a set of curation
procedures, which give additional quality control and automation.
We discuss the methods used and show some example data. Our design is 
particularly effective on correlated data-sets, where the observations 
in different filters are synchronised. It
is scalable to large VISTA datasets which will be observed in the next 
few years and to future surveys such as Pan-STARRS and LSST. 
\end{abstract}

%
%

\vspace{-10mm}
\section{Introduction}
The study of photometrically or astrometrically varying sources has led 
to many important discoveries in astronomy, e.g. stellar masses from 
eclipsing binaries; distance scales from stellar parallaxes and pulsating 
stars; the theory of gravity from planetary motions and the physics of 
accretion disks from observations of cataclysmic variables.
 
With new wide field imagers on survey telescopes, large surveys of variable 
objects have become possible. Over the next few years all sky variability 
surveys will start with Pan-STARRS (Kaiser 2007) and LSST (Walker 2003). 
Near infra-red technology has also improved and the UK Infra-red 
Telescope Wide Field Camera (UKIRT-WFCAM) and the Visible and Infra-red 
Survey Telescope for Astronomy (VISTA; Emerson et al. 2004) 
are the first near infrared cameras capable of undertaking 
large synoptic surveys. Here we discuss the design and implementation of a 
dynamical relational database science archive for archiving synoptic data
observed by UKIRT-WFCAM - WFCAM Science Archive (WSA; Hambly et al. 2008) 
- and by VISTA - VISTA Science Archive (VSA). 

\section{Database Design}

\begin{figure}
\epsscale{.8}
\plottwo{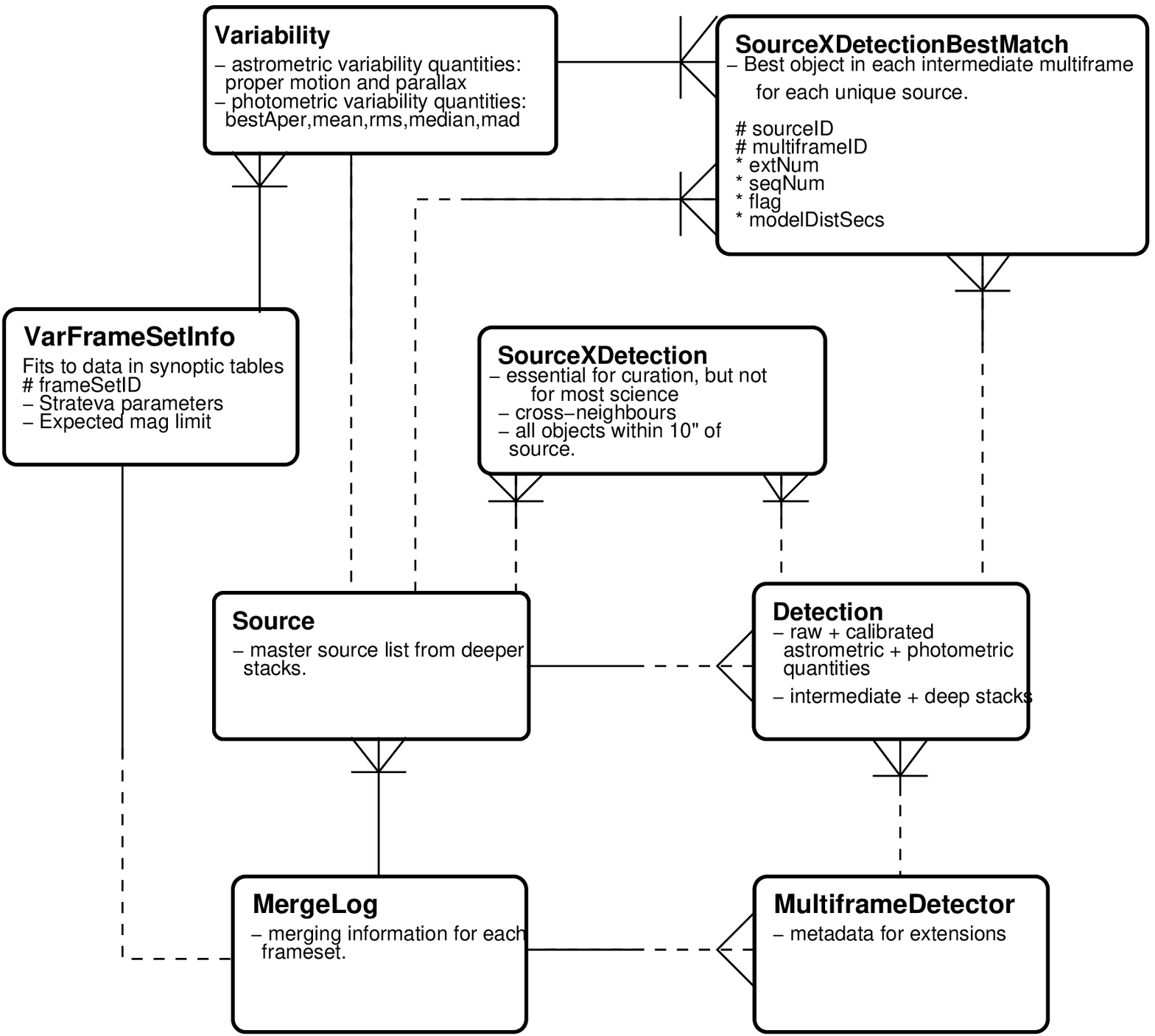}{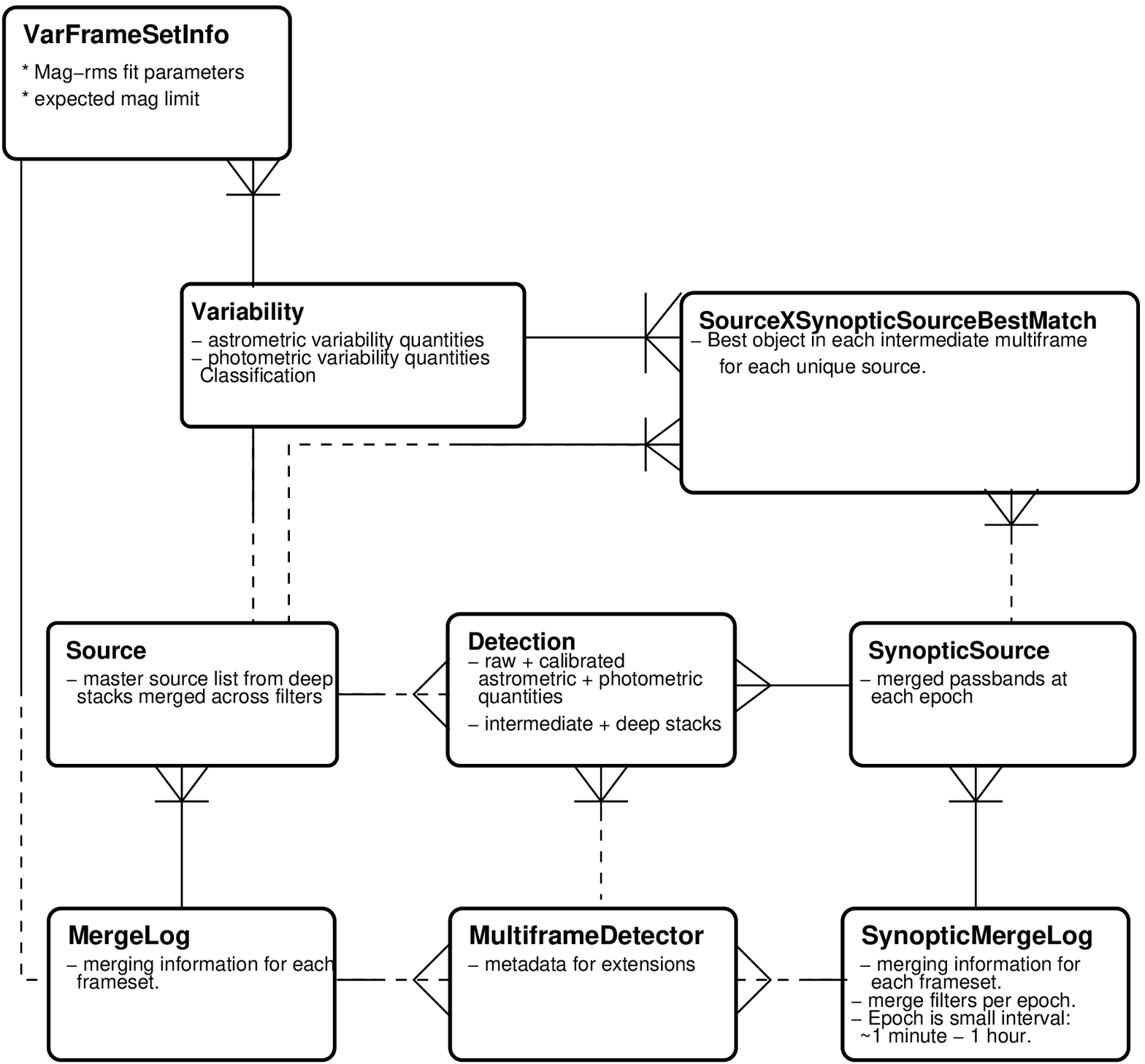}
\caption{\it The left hand plot shows the entity-relation model for 
single passband or uncorrelated multi-epoch data and the right hand plot shows the entity-relation model for correlated multi-epoch data.}
\label{fig:synopticERM}
\end{figure}

The basic design of the WSA and VSA is described in detail in Hambly et al. 
(2008). This work is an extension to incorporate new tables related to using 
multi-epoch data and is an improvement to our initial design (Cross et al. 
2006). Figure ~\ref{fig:synopticERM} shows the new tables and their 
relationship to existing tables in two entity-relation models (ERMs). The 
left-hand ERM is for single or uncorrelated multi-wavelength observations 
and the right-hand side is for correlated multi-wavelength observations.

There are five new tables in the new schema. Two of them only appear in the
correlated multi-epoch ERM and the other three appear in both ERMs. The new 
tables are:

\begin{itemize}

\item The {\bf SynopticMergeLog} table links the different bandpass frames 
that are taken within a short time of each other in a correlated 
multi-epoch survey. It is much like the {\bf MergeLog}, but has the 
additional primary key attribute of {\it meanMjdObs}.
\item The {\bf SynopticSource} table is the merged source catalogue created 
by merging the detection catalogues from the frames in {\bf SynopticMergeLog}. 
This has similar attributes to the {\bf Source} table, but with a primary 
key similar to the {\bf Detection} table. This table makes it easier to 
get colour information at each epoch and makes the {\bf BestMatch} table 
smaller.
\item The {\bf BestMatch} table links the {\bf Source} table - the unique 
list of merged bandpass sources from deep stacks - to each observation: 
either in {\bf SynopticSource} for a correlated bandpass data set or in 
{\bf Detection} for an uncorrelated or single passband data set. 
\item The {\bf Variability} table contains the statistical analysis of the 
multi-epoch data for each source in  the {\bf BestMatch} table. This includes
both astrometric and photometric analysis and classifications.
\item The {\bf VarFrameSetInfo} table contains the fits for data across a whole
frameSet (as defined in {\bf MergeLog}). This is useful for understanding the
limits of the data in each frame set. The model for the astrometric fit is 
also recorded here.
\end{itemize}

These tables can be used together to find and categorise objects. The 
{\bf Variability} table can be used to find objects which have interesting 
statistical properties. Linking this to the {\bf Source} table and to 
neighbour tables of external surveys can select objects with very specific 
morphologies, variable properties, and colours. Only with very
specific selections can a scientist hope to find useful targets for follow up
in a database of $10^9$ objects. The {\bf VarFrameSetInfo} table can help 
define the noise properties of each frameset in each selection. 
The {\bf BestMatch} can then be used with either the 
{\bf Detection} or the {\bf SynopticSource} to display the light curve of
an object. The attributes in each of these tables are described in detail in 
the WSA Schema Browser\footnote{http://surveys.roe.ac.uk/wsa/www/wsa$\_$browser.html}

\section{Archive Curation}

Once new data has been ingested and quality controlled, curation proceeds 
as described in Hambly et al. (2008) and Collins et al. (2009):

\begin{itemize}
\item production of deep stacks and catalogues.  
\item deep stacks in different filters merged into {\bf Source} table.
\item intermediate stacks are recalibrated against deep stacks
\item creation of {\bf SynopticSource} and {\bf SynopticMergeLog}
\item neighbour tables created
\item creation of {\bf BestMatch}
\item creation of {\bf Variability} and {\bf VarFrameSetInfo}
\end{itemize}

These steps are controlled at each stage by the properties in four curation
tables: {\bf RequiredStack}, {\bf RequiredFilters},{\bf RequiredNeighbours} 
and {\bf RequiredSynoptic}. The first three of these are discussed in Collins 
et al. (2009). {\bf RequiredSynoptic} specifies whether a survey has 
correlated passbands and the correlation timescale for the 
{\bf SynopticMergeLog}.  

The new steps that have been added in are the recalibration of 
intermediate stacks that typically improves the zeropoints by $0.005$ mag 
and the synoptic table creation. The recalibration is done by comparing bright 
stars in each intermediate stacks with ones in the deep stack. Frames which 
show big changes in zeropoint are deprecated at this stage. 

The matching of intermediate stacks to each unique source is the most 
important step. The {\bf SourceXDetection} neighbour table
is the starting point for the creation of the {\bf BestMatch} table. The 
neighbour table links the source to all detections within a set radius. 
The {\bf BestMatch} links sources to the nearest match within a smaller 
radius. If there is no match, and the position lies within an observed frame, 
then a default row is inserted. Flags are used to alert users to duplicate 
matches or non-detections close to the edge of a frame. The {\bf BestMatch} 
table is a link between sources and observations, and allows users to track
objects that disappear (an eclipse, or fading after an outburst).

The statistics in the {\bf Variability} table are based on good matches in
the {\bf BestMatch} table. The details of all these steps are in Cross et al. 
(2009). At each stage there are tight constraints on speed, since surveys
such as the VISTA Variables in Via Lactea (VVV) 
survey\footnote{http://www2.astro.puc.cl/VVV/index.php/} - a synoptic survey
of the galactic plane and bulge - 
will have $\sim10^9$ sources, each with 100 observations.

\section{Examples}

\begin{figure}
\epsscale{.8}
\plottwo{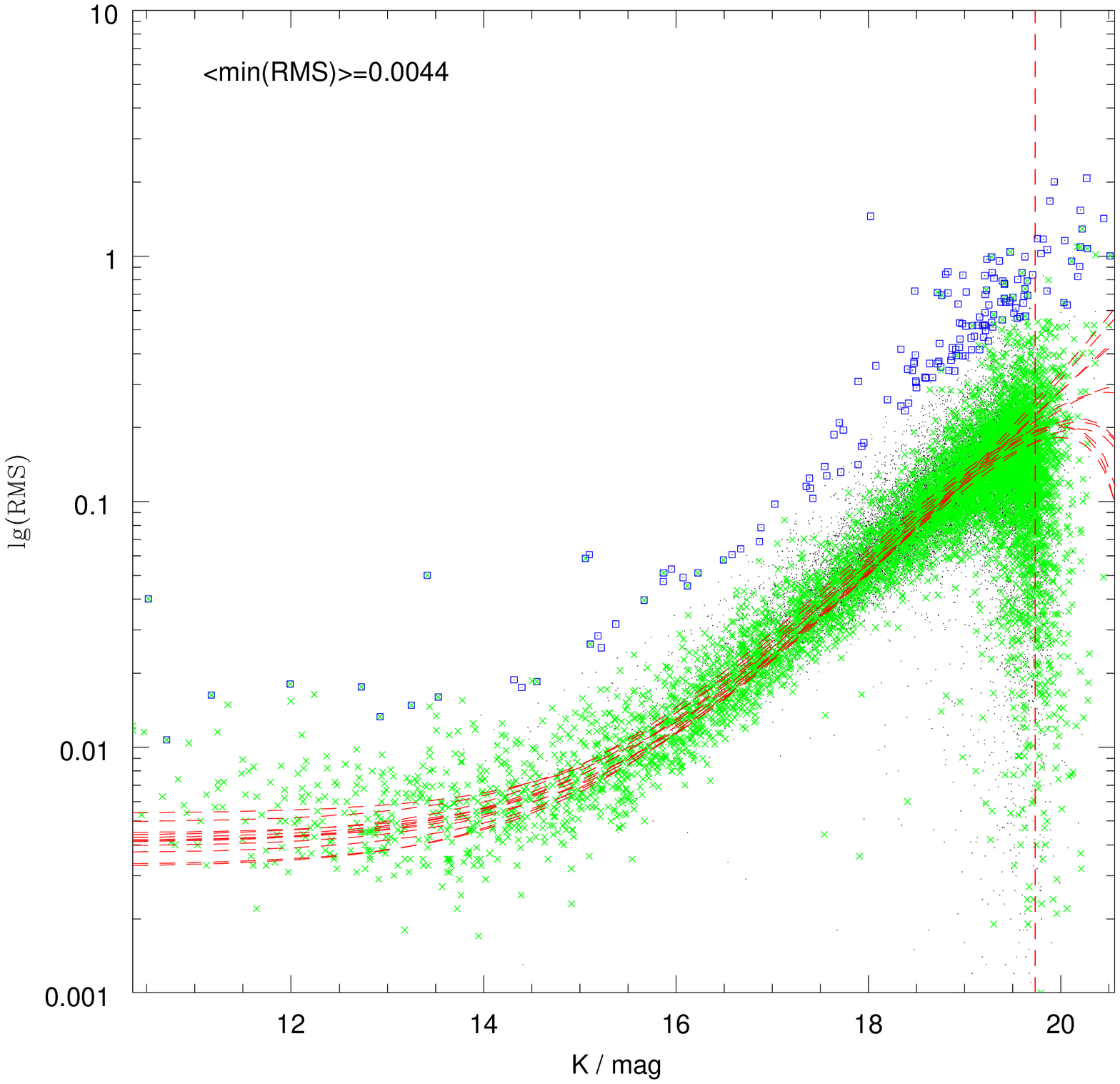}{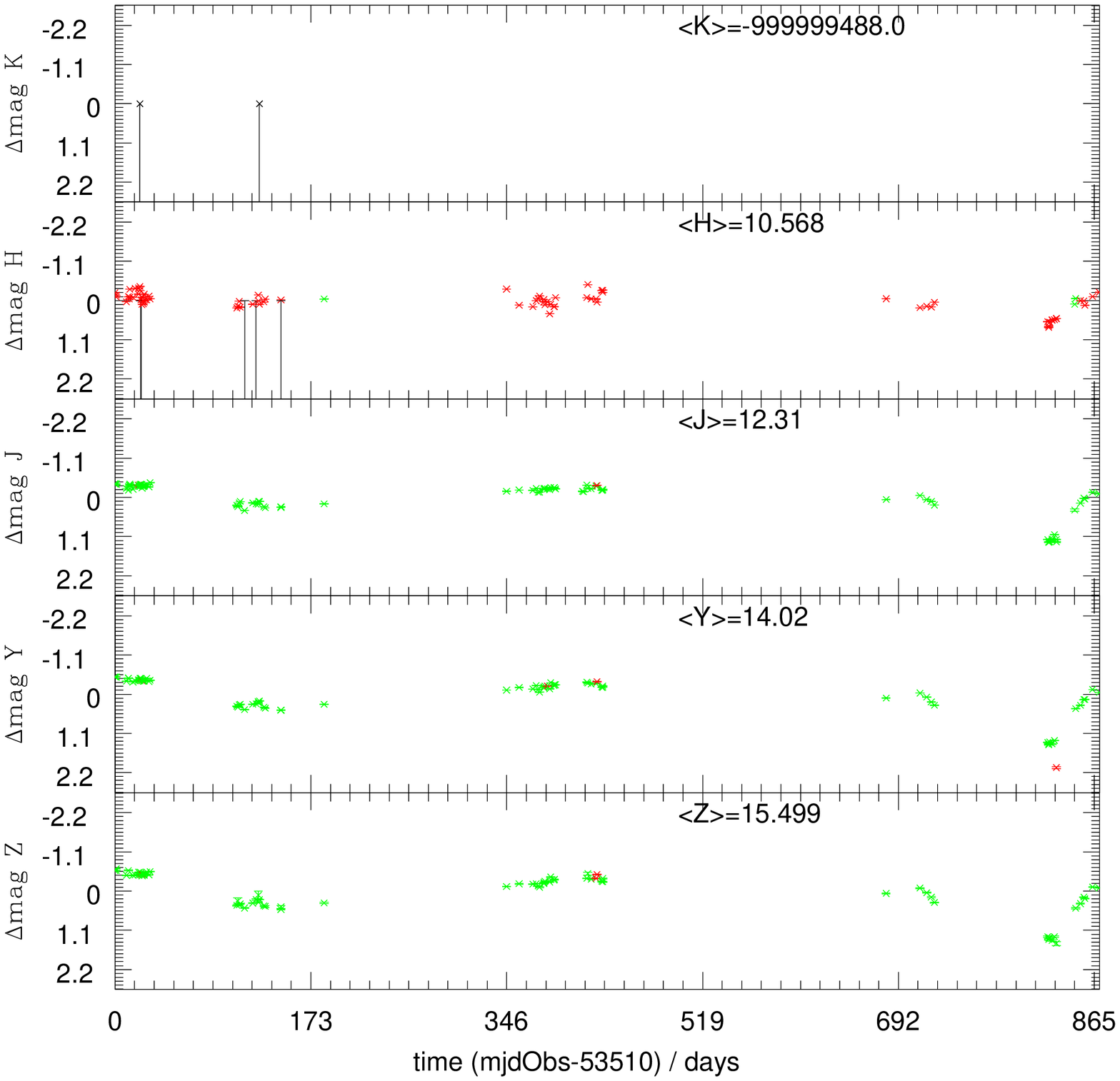}
\caption{\it The left hand plot shows the K-band magnitude-RMS plot for a 
subset of the DXS data. The green crosses are stars and the black 
dots are galaxies. The blue squares are objects classified as variable. The
red curves are fits to the main locus of the magnitude-RMS plot, which give 
the noise properties of the data. The right 
hand plot shows light-curves in all 5 broad-band filters for a single star in
a correlated data set (CAL). Green and red  crosses are good and 
flagged data respectively and black lines are missing data. }
\label{fig:examples}
\end{figure}

We show some example data from two large data sets, the UKIDSS 
(Dye et al. 2006) Deep Extragalactic Survey (DXS) and the UKIRT Standard 
Star calibration data (CAL). Fig~\ref{fig:examples} shows 
an example magnitude-RMS plot (useful for determining which objects are 
variables) and a set of correlated light-curves for a single star. This star
was selected in the archive for its variable characteristics from the WSA.
These plots can be produced using simple SQL queries that are 
described in the WSA SQL cookbook\footnote{http://surveys.roe.ac.uk/wsa/sqlcookbook.html$\#$Multi-epoch}.
 
\begin{references}

\reference Collins, R.\ S.\ et al. 2009, in ASP Conf Ser. XXX, ADASS XVIII, 
ed. D.\ A.\ Bohlender, D.\ Durand \& P.\ Dowler (San Fransisco: ASP),[A.09]
\reference Cross, N.\ J.\ G.\ et al. 2009, in preparation
\reference Cross, N.\ J.\ G.\ et al. 2007, \adassxvi, p54
\reference Dye, S.\ et al. 2006, MNRAS, 372, 1227
\reference Emerson J.\ P.\ Sutherland W.J., et al. 2004, ESO Messenger, 117, 27  
\reference Hambly, N.\ C.\ et al. 2008, MNRAS, 384, 637
\reference Kaiser N.\ 2007, "Proceedings of the Advanced Maui 
Optical and Space Surveillance Conference", ed. S. Ryan, The Maui Economic 
Development Board, p9 
\reference Walker A.\ 2003, MmSAI, 74, 999 
\end{references}
\end{document}